\documentclass[10pt, conference, compsocconf]{IEEEtran}
\usepackage{hyperref}
\hypersetup{
    colorlinks=true,
    linkcolor=black,
    anchorcolor=black,
    citecolor=black,
    filecolor=black,
    menucolor=black,
    runcolor=black,
    urlcolor=black,
    pdftitle={A Partial Reproduction of A Guided Genetic Algorithm for Automated Crash Reproduction},
    pdfpagemode=FullScreen,
    }
    
\usepackage{multirow}

\ifCLASSOPTIONcompsoc
  \usepackage[nocompress]{cite}
\else
  \usepackage{cite}
\fi
\ifCLASSINFOpdf
\else
\fi
\begin{document}
\title{A Partial Reproduction of A Guided Genetic Algorithm for Automated Crash Reproduction}

%% affiliations


% compsoc conferences compared to traditional conferences), use this

\author{\IEEEauthorblockN{Philip Oliver,
Michael Homer,
Jens Dietrich, 
and Craig Anslow}
\IEEEauthorblockA{School of Engineering and Computer Science\\
Victoria University of Wellington\\
Wellington, New Zealand\\ Email: \{philip.oliver, michael.homer, jens.dietrich, craig.anslow\}@vuw.ac.nz}
}


\maketitle

\begin{abstract}
This paper is a partial reproduction of work by Soltani \emph{et al.} which presented EvoCrash, a tool for replicating software failures in Java by reproducing stack traces.
EvoCrash uses a guided genetic algorithm to generate JUnit test cases capable of reproducing failures more reliably than existing coverage-based solutions.
In this paper, we present the findings of our reproduction of the initial study exploring the effectiveness of EvoCrash and comparison to three existing solutions: STAR, JCHARMING, and MuCrash.
We further explored the capabilities of EvoCrash on different programs to check for selection bias.
We found that we can reproduce the crashes covered by EvoCrash in the original study while reproducing two additional crashes not reported as reproduced.
We also find that EvoCrash was unsuccessful in reproducing several crashes from the JCHARMING paper, which were excluded from the original study.
Both EvoCrash and JCHARMING could reproduce 73\% of the crashes from the JCHARMING paper.
We found that there was potentially some selection bias in the dataset for EvoCrash.
We also found that some crashes had been reported as non-reproducible even when EvoCrash could reproduce them.
We suggest this may be due to EvoCrash becoming stuck in a local optimum.
\end{abstract}


\begin{IEEEkeywords}
Automated crash reproduction, empirical software engineering, genetic algorithms, reproduction, search-based software testing.
\end{IEEEkeywords}

% \fi
\IEEEpeerreviewmaketitle

\vspace{-1em}
\section{Introduction}

When software failures occur, developers must manually investigate stack traces and other post-crash information to understand and then replicate the behaviour.
Several tools aim to automate reproducing crashes;
Tools such as STAR, JCHARMING, and MuCrash leverage information produced from a crash to create new unit tests to reproduce the crashes~\cite{chen2015, nayrolles2015, xuan2015}.
However, there are issues with these tools:
STAR cannot handle cases that have external environment dependencies and is affected by the path explosion problem~\cite{chen2015};
MuCrash mutates an existing test suite, so has some reliance on existing tests exploring method sequences of interest~\cite{xuan2015};
and JCHARMING applies computationally expensive model checking~\cite{nayrolles2015}.

Soltani \emph{et al.} presented EvoCrash, a tool using an evolutionary approach that leverages a stack trace to reduce the search space~\cite{soltani2017}.
EvoCrash\footnote{\url{http://www.evocrash.com}} uses the automatic test generation tool, EvoSuite\footnote{\url{http://www.evosuite.org}}, to generate tests.
EvoCrash is an altered version of EvoSuite, which incorporates a novel fitness function developed by Soltani \emph{et al.}
This fitness function is a piece-wise function that checks: the target line number is reached, the correct exception is thrown, and the generated stack trace is similar enough to the original trace~\cite{soltani2016}.
The function is a measure of error and gives a value of 0 when the stack traces match.
This fitness function is used in a guided genetic algorithm to generate tests to replicate stack traces from software crashes.

The guided genetic algorithm uses three genetic operators developed by the original authors.
The first generates an initial population of tests, while the remaining two are altered crossover and mutation operations.
These operators ensure a call to a method within the stack trace contained in each unit test in the search population.

We looked to evaluate the effectiveness of EvoCrash based upon the original paper presented by Soltani~\emph{et al.}~\cite{soltani2017}.
We further extended the suite of crashes used for evaluation to check for selection bias.
Finally, we present some evaluation of discrepancies in the results.
A package containing the supporting data from the original study and our experiments can be found at \texttt{\url{https://doi.org/10.5281/zenodo.5139193}}.

\section{Original Study}

The authors of the original paper~\cite{soltani2017} use EvoCrash to conduct an empirical study with the following two Original Research Questions:
\begin{itemize}
\item{\textbf{ORQ}\textsubscript{1}:} In which cases can EvoCrash successfully reproduce the targeted crashes, and under what circumstances does it fail to do so?

\item{\textbf{ORQ}\textsubscript{2}:} How does EvoCrash perform compared to state-of-the-art reproduction approaches based on stack traces?
\end{itemize}

The initial study was conducted over 50 bugs from \texttt{Apache Commons Collections}\footnote{\url{https://commons.apache.org/proper/commons-collections/}} (ACC), \texttt{Apache Ant}\footnote{\url{http://ant.apache.org}} (ANT), and \texttt{Apache Log4j}\footnote{\url{http://logging.apache.org/log4j/1.2}} (LOG)~\cite{soltani2017}.
The generation of tests for each bug was repeated 50 times to account for the random nature of the guided genetic algorithm.
Soltani \emph{et al.} selected widely used parameter values for the evolutionary component of EvoCrash:
\begin{itemize}
\item{\textbf{Population size:}} Initially set to 50, but increased by 25 iteratively up to 300 if the fitness value does not reach 0.0.
\item{\textbf{Crossover probability:}} Set to 0.75.
\item{\textbf{Mutation probability:}} Set to \texttt{1/n}, with \texttt{n} being the length of the current test case.
\item{\textbf{Search timeout:}} Set to 30 minutes with early stopping if the fitness value reaches 0.0.
\end{itemize}

Regarding the mutation probability, we cross-referenced the paper referenced in the original study.
Fraser and Arcuri state the mutation probability in EvoSuite is \texttt{1/n}, with \texttt{n} being the size of the \emph{test suite}~\cite{fraser2013}.
This probability results in one test case in the suite being mutated on average, rather than one statement in a test case being altered, as reported by Soltani \emph{et al.}
It is not clear if Soltani \emph{et al.} have altered the mutation probability as such in EvoCrash.

Soltani \emph{et al.} selected two metrics for \textbf{ORQ}\textsubscript{1} proposed by Chen and Kim~\cite{chen2015}.
\emph{Crash Coverage} ensures that the crash has been successfully replicated by comparing the exception type thrown and the source line from which it is thrown.
The original authors consider a crash to be covered when a fitness value of 0.0 is reached.
\emph{Test Case Usefulness} concludes that a test case is useful if it reveals the bug which caused the crash.
Two of the original authors independently performed manual validation to decide if the test cases produced by EvoCrash successfully reveal the bug.
In the case of disagreements, the conclusions were discussed.
We do not assess the test case usefulness as a part of our reproduction.
This omission is primarily due to the subjective nature of this metric.

\textbf{ORQ}\textsubscript{2} was investigated through comparison with three other crash reproduction technologies: STAR, MuCrash, and JCHARMING.
Soltani \emph{et al.} used published data from these tools, as the artifacts for the tools were unavailable at the time of writing.
The comparison to STAR was completed using 50 of the 52 bugs collected by Chen and Kim~\cite{chen2015}.
EvoCrash was compared to JCHARMING using 8 of the 20 bugs collected by Nayrolls~\emph{et al.}~\cite{nayrolles2015}.
Finally, the comparison to MuCrash was performed using the 12 ACC bugs collected for testing STAR \emph{et al.}~\cite{xuan2015}.
Several bugs were excluded from the original study.

\textbf{EvoCrash Performance ORQ\textsubscript{1}}.
The original paper presents results for \textbf{ORQ}\textsubscript{1}, with EvoCrash successfully replicating 41 of the 50 (82\%) bugs~\cite{soltani2017}.
EvoCrash reproduced 10 out of 12 bugs for ACC, 14 out of 20 for ANT, and 17 out of 18 for LOG.
EvoCrash does not support the six unreproducible cases for ANT due to dependencies on missing external \texttt{build.xml} files.
One of the cases from LOG is unsupported due to a call to a static class initialiser.
The two unreproducible cases for ACC are due to the complexity of the bugs.
Using the \emph{Test Case Usefulness} criteria from Chen and Kim, the original authors conclude that 34 of the 39 generated test cases were useful.
The other 5 test cases mainly were found to have dependencies on external files, which were not available.
In this study, we do not explore the usefulness of the test cases generated by EvoCrash.

It is unclear what threshold the original authors have used to discern whether a bug has been replicated.
They state that ``of the replicated cases, the crash LOG-509 had the lowest rate of replications - 39 out of 50,'' with these numbers being 39 replications of the crash over 50 runs~\cite{soltani2017}.
However, they also state that for one of the non-reproducible cases (ACC-104), ``EvoCrash could replicate the case 4 times out of 50.''
While this is a complex bug that requires a specific order of method calls to trigger the crash, it would appear that EvoCrash can successfully replicate the behaviour, albeit occasionally.

\textbf{Comparison to Other Tools ORQ\textsubscript{2}}.
Compared with STAR, EvoCrash has almost identical results, except for ACC-104 (discussed above)~\cite{soltani2017}.
EvoCrash is also capable of replicating three additional cases which are prone to the path explosion problem.
Compared with MuCrash, EvoCrash can replicate all the crashes replicated by MuCrash and an additional 3 cases, with one of these cases marked as not useful.
EvoCrash covers all the crashes successfully reproduced by JCHARMING (6 out of 8) and can reproduce the two crashes JCHARMING cannot.
However, 3 of the test cases from EvoCrash are marked as not useful, with two being crashes JCHARMING could reproduce.
Nayrolles \emph{et al.} do not identify if crashes reproduced by JCHARMING are useful; therefore, it could be that the non-useful tests generated by EvoCrash are also not useful when generated with JCHARMING~\cite{nayrolles2015}.

\vspace{-0.3em}
\section{Reproduction}

For the reproduction in this study, we performed two experiments using the publicly available reproduction package\footnote{\url{https://github.com/STAMP-project/EvoCrash/releases/tag/evocrash-refactored}} for the original paper.
The first experiment was run using the parameters and configuration as-is from the package.
Following the further investigation into the parameters used in the package, we found that some did not match what was reported in the original paper for population sizes.
Many population sizes were initialised at 80, which does not follow the experiment procedure outlined in the initial study.
We increased these to the next largest population size that fit the procedure for population sizes that did not conform to the experimental procedure.
The experiment was rerun using these updated parameters.
In the second experiment, we followed the initial study's guidelines to increase the population sizes by 25 repeatedly up to 300 for crashes which cannot be reproduced.
All other parameters used match the experiment procedure from the initial study.

There were a few issues when beginning the reproduction.
Firstly, the website for the package location in the original paper no longer exists.
This issue was circumvented by finding the publicly available release package on GitHub.
The second issue was that the scripts used to run EvoCrash for the 50 crashes were not OS-agnostic.
Classpath separators had been hardcoded as semicolons (;) for use on a Windows machine.
These separators were changed to run the experiment on Arch linux successfully.
Thirdly, some of the paths for the binaries for the targeted programs were incorrect.
For example, there were a few cases of the \texttt{LOG4jb-1.0.4/} directory being referenced as \texttt{Log4jb-1.0.4/}.
These paths were fixed for the experiment.
Another issue was that some of the results from the original study were missing from the reproduction package.
The 30\textsuperscript{th} run is missing most of the results, while some other runs do not have the results for some crashes.
Finally, \texttt{ACC-377} was missing from the crashes and results in the reproduction package.
This crash was added to the experiment to ensure similarity between the original experiment and the reproduction.

After replicating the main results from the study, we looked to evaluate EvoCrash on some other crashes, including those from the STAR and JCHARMING papers which were excluded from the original study.
We also selected 7 crashes from \texttt{Apache Commons Lang}\footnote{\url{https://commons.apache.org/proper/commons-lang/}} (ACL) and 6 crashes from \texttt{Apache Commons BeanUtils}\footnote{\url{https://commons.apache.org/proper/commons-beanutils/}} (BEAN) to check for selection bias in the initial dataset.

\begin{table}[tb]
\centering
\caption{Results from original paper and reproduction. Percentages of 100\% are not reported for brevity}
\label{T:reproducedCrashes}
\begin{tabular}{ l l l l l }
\hline
Project & Bug ID & Original & Experiment 1 & Experiment 2 \\
\hline
\multirow{12}{4em}{ACC} & 4 & Y & Y & Y \\
& 28 & Y & Y & Y \\
& 35 & Y & Y & Y \\
& 48 & Y & Y & Y \\
& 53 & Y & Y & Y \\
& 68 & N (0\%) & N (0\%) & N (0\%) \\
& 70 & Y & Y (100\%) & Y (98\%) \\
& 77 & Y & Y& Y\\
& \textbf{104} & \textbf{N (8\%)} & \textbf{Y (2\%)} & \textbf{Y (8\%)} \\
& 331 & Y (82\%) & Y (52\%) & Y (88\%) \\
& 377 & Y & Y (90\%) & Y (60\%) \\
& 441 & Y & Y & Y \\
\hline
\multirow{20}{4em}{ANT} & 28820 & N (0\%) & N (0\%) & N (0\%) \\
& 33446 & Y & Y & Y \\
& 34722 & Y & Y & Y \\
& 34734 & Y & Y & Y \\
& 36733 & Y & Y & Y \\
& 38458 & Y (92\%) & Y (90\%) & Y (90\%) \\
& 38622 & Y (80\%) & Y (86\%) & Y (82\%) \\
& 42179 & Y & Y & Y \\
& \textbf{43292} & \textbf{N (94\%)} & \textbf{Y (96\%)} & \textbf{Y} \\ % original report has 2 failures, 47 successes, and 1 unreported (for run #30)
& 44689 & Y & Y & Y \\
& 44790 & Y & Y & Y \\
& 46747 & N (0\%) & N (0\%) & N (0\%) \\
& 47306 & N (0\%) & N (0\%) & N (0\%) \\
& 48715 & N (0\%) & N (0\%) & N (0\%) \\
& 49137 & Y & Y & Y \\
& 49755 & Y (94\%) & Y & Y \\
& 49803 & Y & Y & Y (98\%) \\
& 50894 & Y & Y & Y \\
& 51035 & N (0\%) & N (0\%) & N (0\%) \\
& 53626 & Y & Y & Y \\
\hline
\multirow{18}{4em}{LOG} & 29 & Y (88\%) & Y (90\%) & Y (96\%) \\
& 43 & N (0\%) & N (0\%) & N (0\%) \\
& 509 & Y (74\%) & Y (50\%) & Y (78\%) \\
& 10528 & Y & Y & Y \\
& 10706 & Y & Y & Y \\
& 11570 & Y & Y & Y \\
& 31003 & Y & Y & Y \\
& 40212 & Y & Y & Y \\
& 41186 & Y & Y & Y \\
& 44032 & Y & Y & Y \\
& 44899 & Y & Y & Y \\
& 45335 & Y (94\%) & Y (94\%) & Y (96\%) \\
& 46144 & Y (82\%) & Y (78\%) & Y (86\%) \\
& 46271 & Y (94\%) & Y & Y \\
& 46404 & Y & Y & Y \\
& 47547 & Y & Y & Y \\
& 47912 & Y & Y & Y \\
& 47957 & Y & Y & Y \\
\hline \\
\end{tabular}
\\ 
Y - Crash has been replicated at least once \\
N - Crash has not been replicated \\
Percentage values are the number of successful replications from 50 runs
\vspace{-2em}
\end{table}

\vspace{-0.3em}
\subsection{Experimental Results}

Table~\ref{T:reproducedCrashes} presents the original study's results alongside the results we have achieved over our two runs of the experiment.
It can be seen that our results are mainly similar to those in the original study, with two notable exceptions: \texttt{ACC-104} and \texttt{ANT-43292}.
As previously discussed, \texttt{ACC-104} is successfully reproduced by EvoCrash in the original study, albeit at a rate of 8\%.
In our experimental runs, we achieved success rates of 2\% and 8\%.
The original authors were looking to answer the research question of whether EvoCrash could reproduce a crash.
We argue that even a single success means EvoCrash can reproduce the crash.
We further argue that a low reproduction rate could indicate issues within the initialisation of the genetic programming parameters.
It could be possible that EvoCrash becomes stuck in a local optimum with not enough mutation occurring to allow the program to find a better test case.

In the case of \texttt{ANT-43292}, the original study marked this crash as not reproduced.
We found 96\% and 100\% success rates for this crash in our experiments.
On closer inspection of the data from the original study, we found that the crash was successfully reproduced.
In the underlying data, we found 47 successful reproductions, with two failures and one unreported result.
This data gives a success rate of 94\% for \texttt{ANT-43292} in the original study.
It could be that the original authors meant to mark this crash as \emph{not useful}.
However, we do not confirm that this is the case.

\begin{table}[tb]
\centering
\caption{Results from crashes excluded from original study, including comparison to STAR and JCHARMING}
\label{T:excludedCrashes}
\begin{tabular}{ l l l c c }
\hline
Project & Bug ID & Results & STAR & JCHARMING \\
\hline
\multirow{1}{4em}{DnsJava} & 38 & N (0\%) & - & Y \\
\hline
\multirow{3}{4em}{Jfreechart}& 434 & Y (98\%) & - & Y \\
& 664 & N (0\%) & - & Partial \\
& 916 & N (0\%) & - & Y \\
\hline
\multirow{2}{4em}{Pdfbox} & 1359 & N (0\%) & - & N \\
& 1412 & Y (94\%) & - & Partial \\
\hline
\multirow{1}{4em}{ANT} & 41422 & Y (100\%) & Y & N \\
\hline \\
\end{tabular}
\\
Y - Crash has been replicated at least once \\
N - Crash has not been replicated \\
Percentage values are the number of successful replications from 50 runs
\vspace{-2em}
\end{table}

Table~\ref{T:excludedCrashes} presents the results of crashes from \texttt{DnsJava}\footnote{\url{https://github.com/dnsjava/dnsjava}}, \texttt{Jfreechart}\footnote{\url{https://www.jfree.org/jfreechart/}}, \texttt{Pdfbox}\footnote{\url{https://pdfbox.apache.org/}}, and \texttt{ANT} which were used in the STAR and JCHARMING papers~\cite{chen2015,nayrolles2015}.
In the JCHARMING paper there were also crashes used from \texttt{ArgoUML}\footnote{\url{https://github.com/argouml-tigris-org}} and \texttt{Open Mission Control Software}\footnote{\url{https://nasa.github.io/openmct/}}~\cite{nayrolles2015}.
However, we could not find the stack traces for these crashes and thus have not included them.
% We also selected some crashes from \texttt{Apache Commons Lang} and \texttt{Apache Commons BeanUtils} to search for selection bias and to discover how well EvoCrash performs on more difficult stack traces.
The crashes excluded from the original study do not have a high success rate of reproduction by EvoCrash, with 4 of the 13 crashes reproduced (43\%).
If these crashes were included in the original study, EvoCrash would have reproduced 44 out of 57 crashes (77\%), rather than the 82\% reported~\cite{soltani2017}.

Table~\ref{T:excludedCrashes} also shows the comparison of the crashes excluded from the original study to STAR and JCHARMING.
The main comparisons here are between EvoCrash and JCHARMING for the \texttt{DnsJava}, \texttt{Jfreechart}, \texttt{Pdfbox}, and \texttt{ANT} crashes.
Of these seven crashes, JCHARMING reproduced 5, while EvoCrash reproduced 3.
Of the 15 crashes shared by EvoCrash and JCHARMING, JCHARMING successfully reproduced 11 (73\%), while EvoCrash also reproduced 11 (73\%).
It is of particular interest that JCHARMING is capable of reproducing \texttt{DnsJava-38}, \texttt{Jfreechart-664}, and \texttt{Jfreechart-916} where EvoCrash cannot.
The original study found a significant difference between the performances of EvoCrash and JCHARMING~\cite{soltani2017}.
However, it is clear that with other crashes from the JCHARMING paper, the performance is similar.

Table~\ref{T:novelCrashes} presents the results of our evaluation on crashes from \texttt{Apache Commons Lang} and \texttt{Apache Commons BeanUtils}.
We have selected these crashes to identify any potential for selection bias in the original study.
Of the ACL crashes, EvoCrash successfully reproduced 4 out of 7 (57\%).
The three failing tests use date formats or message formats, which require specifically formatted strings as input.
It is therefore unsurprising that EvoCrash cannot reproduce these crashes, as it has not been created with the capability of consistently generating strings that match the complex formats required by these classes.
Finally, given the complexity of configuration required to use \texttt{BeanUtils} in a program, the 0\% success rate is unsurprising.
As most of these BEAN crashes arise due to configuration issues, EvoCrash struggles to generate a test case to initialise such a configuration.

\vspace{-0.5em}
\section{Discussion}
\vspace{-0.5em}

The authors of the original paper set out to evaluate the tool, EvoCrash, on several crashes and to compare these results with the existing tools: STAR, MuCrash, and JCHARMING~\cite{soltani2017}.
The original study successfully reproduced 41 of 50 (82\%) crashes.
We found that EvoCrash can successfully reproduce all the crashes presented in the original study through our two main experiments.
We also found two crashes (\texttt{ACC-104} and \texttt{ANT-43292}) which we reproduced with EvoCrash, but are not reported as reproduced in the original study.
We found in the data underlying the original study that \texttt{ANT-43292} has a 94\% reproduction rate, while our experiments have 96\% and 100\% reproduction rates.
This misidentified result in the original study likely occurred due to human error.
We consider a crash to be reproduced if it can be successfully reproduced in at least one run.
Crashes with low reproduction rates could point to issues in the genetic parameters for EvoCrash, as there may not be enough variability introduced to allow EvoCrash to escape local optima.

\begin{table}[t]
\centering
\caption{Results from additional crashes used in this study}
\label{T:novelCrashes}
\begin{tabular}{ l l l }
\hline
Project & Bug ID & Results\\
\hline
\multirow{7}{4em}{ACL} & 948 & N (0\%) \\
& 1186 & N (0\%) \\
& 1192 & N (0\%) \\
& 1276 & Y (100\%) \\
& 1292 & Y (100\%) \\
& 1310 & Y (86\%) \\
& 1385 & Y (100\%) \\
\hline
\multirow{6}{4em}{BEAN} & 276 & N (0\%) \\
& 302 & N (0\%) \\
& 351 & N (0\%) \\
& 421 & N (0\%) \\
& 541 & N (0\%) \\
& 547 & N (0\%) \\
\hline \\
\end{tabular}
\\
Y - Crash has been replicated at least once \\
N - Crash has not been replicated \\
Percentage values are the number of successful replications from 50 runs
\vspace{-2em}
\end{table}

We present a comparison between EvoCrash, STAR, and JCHARMING for crashes excluded from the original study.
For \texttt{ANT-41422}, EvoCrash and STAR could both successfully reproduce this crash; however, JCHARMING could not.
For the other crashes from \texttt{DnsJava}, \texttt{Jfreechart}, and \texttt{Pdfbox}, we found that JCHARMING outperforms EvoCrash, contrasting with the original result that EvoCrash outperformed JCHARMING.
We find that EvoCrash and JCHARMING both reproduce 73\% of crashes once the full JCHARMING dataset is used.
This result could potentially point to some selection bias in the original study, as these crashes were excluded.
As JCHARMING, STAR, and MuCrash are not publicly available, selection bias could be present in the dataset chosen for those studies.

While we do not analyse the usefulness of the test cases generated by EvoCrash, we did consider the suitability of the metric for this.
The metric requires that the buggy stack frame exists in the reproduced stack trace.
A number of the crashes reproduced by EvoCrash are attempting to reproduce only one stack frame in a larger stack trace.
A potential question for future work is raised: whether this metric is suitable and if the crashes can be considered reproduced if this metric is met.
Furthermore, this metric is subjective and cannot be easily reproduced.
Comparisons between the crashes reproduced by EvoCrash and the actual bug fixes committed to the source code could be drawn to clarify that the tests generated correctly identify a bug and relate to the bug-fix in the main project.

We conclude that EvoCrash is a tool that can be used to reproduce several crashes in Java successfully.
However, we are not sure the data presented in the original paper is representative of the capabilities of EvoCrash.
Several low-performing crashes appear to have been excluded from the original study, including those which contribute significantly to the original paper's conclusion that EvoCrash performs significantly better than JCHARMING.
We also suggest there may be issues with the parametric setup of the genetic part of EvoCrash, leading to low variability and the system becoming stuck in local optima.
Future work could look into these issues and the usefulness of the test cases produced by EvoCrash.

\bibliographystyle{IEEEtran}
\bibliography{references}

\end{document}